\documentclass[aps, reprint]{revtex4-1}

\usepackage[utf8]{inputenc}
\usepackage{graphicx}
\usepackage{hyperref}
\usepackage{amsmath}

\begin{document}

\title{Computationally efficient Monte Carlo electron transport algorithm for nanostructured thermoelectric material configurations}

\author{Pankaj Priyadarshi}
\email{pankaj.priyadarshi@warwick.ac.uk}
\author{Neophytos Neophytou}
\email{n.neophytou@warwick.ac.uk}

\affiliation{School of Engineering, University of Warwick, Coventry, CV4 7AL, United Kingdom \\}


\begin{abstract}
Monte Carlo statistical ray-tracing methods are commonly employed to simulate carrier transport in nanostructured materials. In the case of a large degree of nanostructuring and under linear response (small driving fields), these simulations tend to be computationally overly expensive due to the difficulty in gathering the required flux statistics. Here, we present a novel MC ray-tracing algorithm with computational efficiency of at least an order of magnitude compared to existing algorithms. Our new method, which is a hybrid of analytical Boltzmann transport equation and Monte Carlo uses a reduced number of ray-tracing particles, avoids current statistical challenges such as the subtraction of two opposite going fluxes, the application of a driving force altogether, and the large simulation time required for low energy carriers. We demonstrate the algorithm's efficiency and power in accurate simulations in large domain nanostructures with multiple defects. We believe that the new method we present is indeed more robust and user friendly compared to common methods, and can enable the efficient study of transport in nanostructured materials under low-field steady-state conditions.
\end{abstract}
\maketitle
\section{Introduction}
\begin{figure*}[t]
\includegraphics[height=0.4\textwidth,width=1\textwidth]{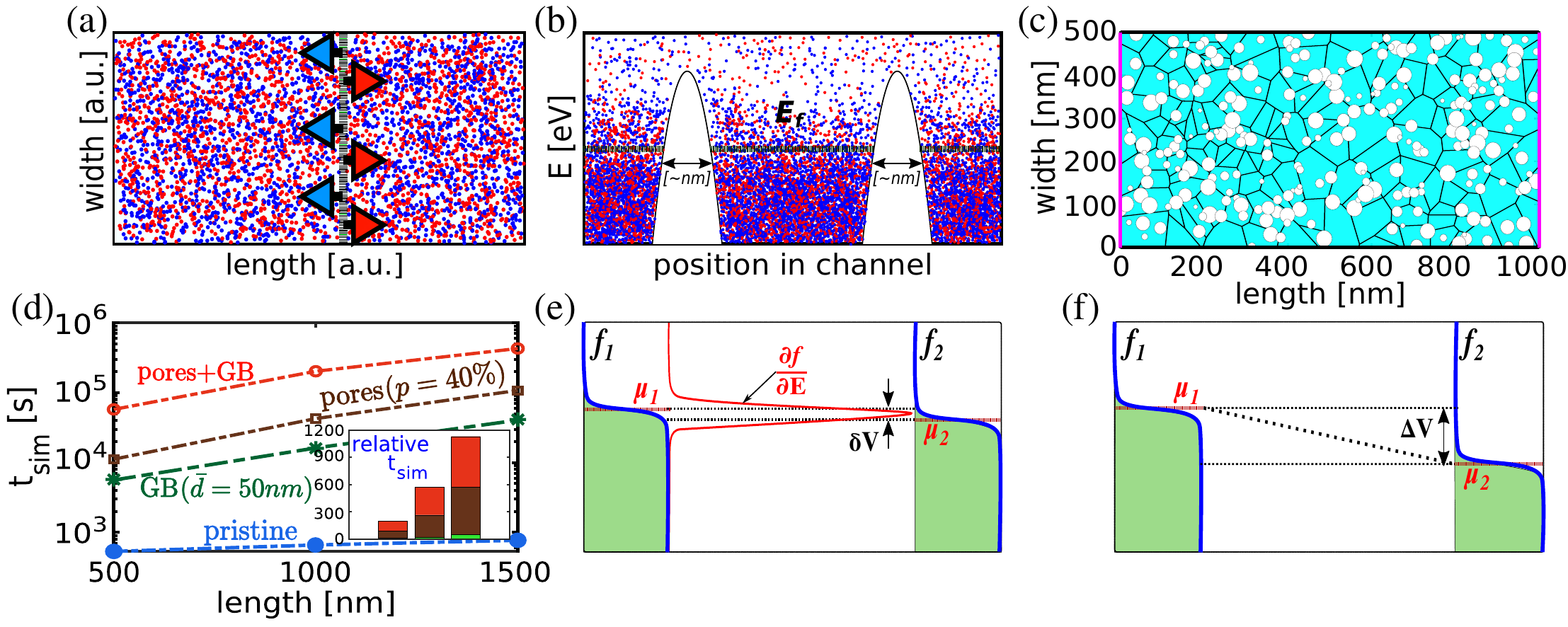}

\caption{(a) Schematic of typical ensemble Monte Carlo particle distribution, with left and right going particles as depicted by blue and red arrows, respectively, in a bulk material, and (b) in a nanostructured material with potential barriers. (c) Grain boundary and nanoporous populated nanostructured domain. (d) Simulation time as a function of channel length of various nano-featured 2D domains (pristine material, grain boundaries - GB in the material, pores in the material, and pores plus GB in the material). The inset plot shows the relative (or ratio of $t_{\text{sim}}$) time of MC simulation in the nanostructured feature domain with respect to the pristine domain. (e) The energy derivative of the Fermi distribution (in red), which is mimicked by a small applied potential $\delta V$ in low-field transport conditions, typically in a micrometer length channel domain. (f) Significantly large $\Delta V$ that splits the contact Fermi levels $\mu_1$ and $\mu_2$, imposing high field transport conditions.}
\label{P01_01}
\end{figure*}

A plethora of new materials and their alloys have recently been synthesized for a variety of applications, and typically most of them are poly- or nano-crystalline with embedded defects and often a large degree of non-uniformities, as for example in thermoelectric (TE) materials\cite{Beretta2019}. Specifically for TEs, these materials are nanostructured, and they are so on purpose. In order to evaluate their transport properties, we typically employ the semi-classical Boltzmann transport equation (BTE). The solution of the BTE can be evaluated either analytically or numerically \cite{LundstromBook2000,JacoboniBook2010,Kosina2003_1,TomizawaBook1993,SelberherrBook2013}. Since it is a seven dimensional integro-differential equation (six dimensions in the phase space and one in time), its analytical solution is cumbersome and can be solved under very restrictive assumptions \cite{GoldsmidBook2010,WangBook2014,FriedmanBook1991}. A grid-based deterministic numerical method such as the renewed spherical harmonics approach is sometimes used, but it requires very powerful computational platforms \cite{HongBook2011,Hong2009,Rupp12016,BanooThesis2000,Vecchi1998,Gnudi1993}. Another way to solve the BTE is based on stochastic solution methods using Monte Carlo (MC) computational algorithms, which use statistical sampling to solve the BTE numerically, and are frequently used for electronic device applications \cite{Jacoboni1983,Fischetti1988, Jacoboni_LugliBook1989,Hess1991,Eastman1997,Borowik2005,Walid2012}. Over the last several years, MC techniques also found extensive use in the fields of charge and energy transport in semiconductor materials \cite{Dhritiman2019,Wolf2014_1,Wolf2014_2,Maurer2015,Fischetti1991,Fischetti2004}. 
MC simulation methods are particularly useful for nanostructured materials, where the carriers encounter a plethora of defects along their transport direction and interact with them, while analytical solutions in these cases are not as accurate. The simplest way to consider theoretically the influence of nanostructuring on carrier transport is to use Matthiessen's rule to add an additional geometry dependent scattering rate on top of the internal bulk material scattering processes in the analytical BTE \cite{Huang2006,TermentzidisBook2014}. However, it is never clear how to determine this geometrical length scale in complex nanostructured materials with multiple and irregularly placed defect types. Using stochastic MC simulations, the transport in nanostructured configurations can be modeled in a real space domain, and proper insight can be gained \cite{Wolf2014_1,LacroixJAP2014,Ziqi2017,Dhritiman2018,DhritimanLaura2019,NeophytouBook2020}. \\
\indent One category of materials which has benefited from nanostructuring over the last two decades is thermoelectric materials, which convert the heat energy from a temperature gradient to electrical energy and vice-versa \cite{HandbookTE1995,Snyder2008,Beretta2019}. Their conversion efficiency is controlled by both their charge and heat/energy transport. New-generation TE materials are typically highly nanostructured, with nano-features spanning from macro- to nano-scale (including boundaries, potential barriers, pores, nanoinclusions, atomic defects, second-phasing, etc.) \cite{Biswas2012,Liu2012,Nakamura2015,Taborda2016,Srinivasan2019,Zheng2021}  . The recent improvement in the performance of TE materials originates in most cases from reduced thermal conductivity due to phonon scattering with the boundaries of the nano-defects, and thus MC simulations are typically performed for phonon transport in real space in such materials \cite{Mazumder_Majumdar_2001,LacroixJAP2014,LacroixPRB2014,Dhritiman2018,Dhritiman2019}. Nevertheless, studies on electronic transport using MC are also emerging for nanostructured TE materials after it was pointed out that specific designs can also improve the so-called power factor, which directly determines the efficiency of a TE material as well \cite{Neophytou_2013,Neophytou_2019}.\\
\indent Monte Carlo simulations involve the ray-tracing of particle trajectories rather than the direct solution of partial differential equations. These particles are allowed to move in the domain in both left/right directions under the influence of a driving force (as shown in Fig.~\ref{P01_01}(a), and the net flux is computed in a statistical manner. Although this method served well simulations in bulk materials with success over many years, for nanostructured materials large difficulties are encountered, which make simulations computationally extremely expensive and logistically cumbersome. The presence of multiple scattering sites from nanostructuring reduces the flux in the domain at such a degree, which makes it very difficult to gather enough statistics for converged flux results. This is particularly difficult under linear response, where the two opposite going fluxes vary only slightly. Typical simulations in the literature require from 10s of thousands to even millions of trajectories for adequate results \cite{VasileskaBook2010,EricPop2004,Jungemann2001,Zebarjadi2006,Zebarjadi2007,Dhritiman2018}. \\
\indent In order to address the above issues and enable efficient and accurate MC simulations for nanostructures, we have developed a hybrid MC algorithm which: i) merges information from analytical BTE solutions with the numerically extracted flux, ii) considers only a single flux initialized from the left only and injected into the channel where it is ray-traced to either of the contacts, iii) does not require the application of a driving force. The method we present provides the same accuracy as common methods, but with a significantly reduced computational cost. \\ 
\indent This paper is organized as follows: In Sec. II we describe the challenges encountered in existing MC algorithms for nanostructures and our method which provides improved computational cost. A stochastic error analysis and convergence is presented in Sec. III. How this method is very efficient and effective for transport in nanostructures is discussed in Sec. IV. Finally, in Sec. V we conclude. \\
\section{Computational Formalism}
\begin{figure}
\centering
\includegraphics[height=0.51\textwidth,width=0.48\textwidth]{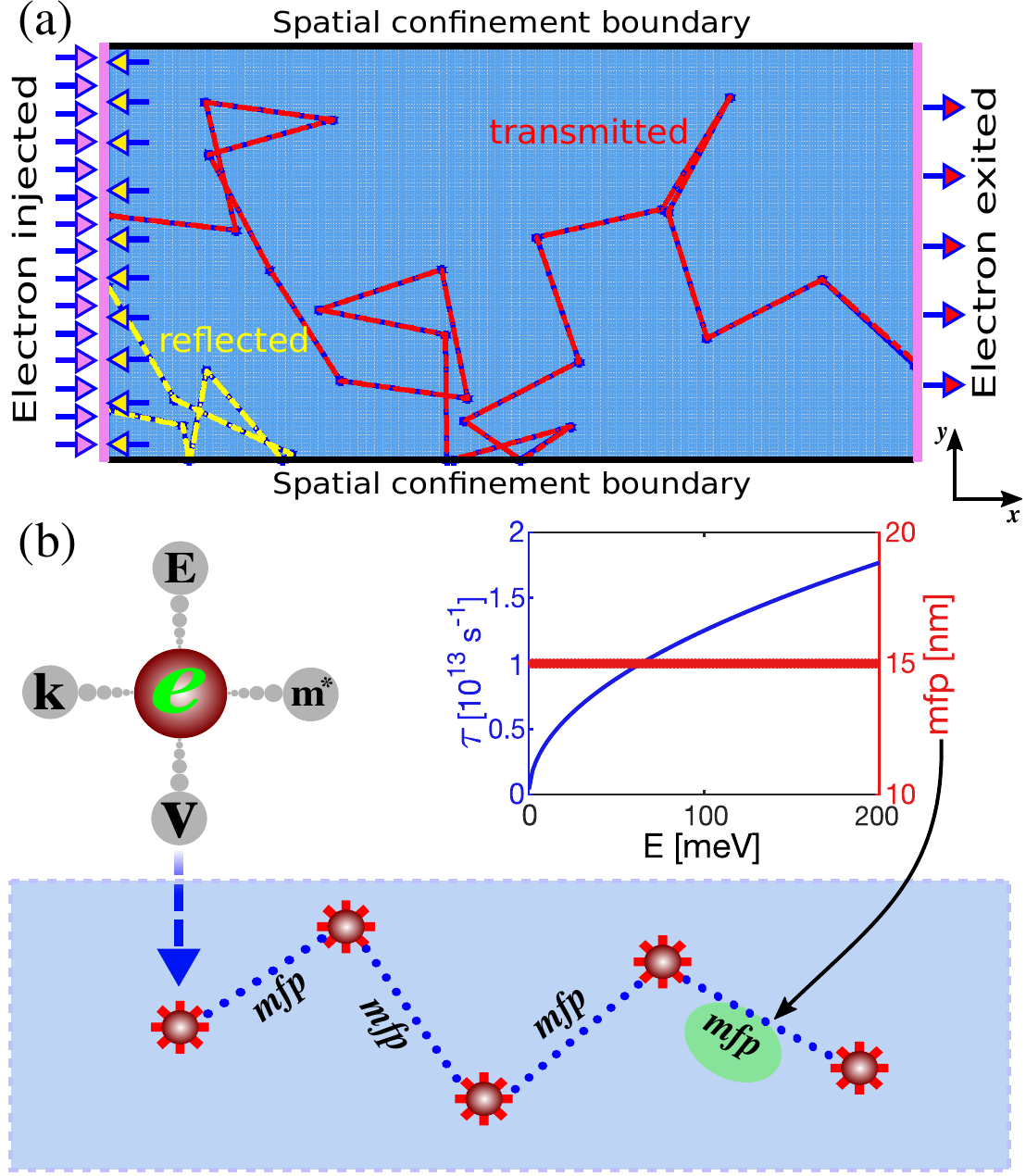}

\caption{(a) A pristine domain having electrons injected from the left, which traverse across the domain with scattering intermediate events after every mean-free-path ($mfp$). Few electrons are transmitted and reached to the right end (as depicted by the red line), but many of them are reflected back to the left end (in yellow color) after multiple $mfp$ steps. The top and bottom ends are spatially firmed so that electrons are only allowed to specularly reflect from the boundaries. (b) The parameters that characterize an electron in the Monte Carlo algorithm, dictated by the assumed band dispersion and the propagation direction, i.e. energy $E$, momentum $k$, conductive effective mass $m^*$, and velocity $v$. A free flight distance between two consecutive collisions i.e. the mean-free-path $mfp$, is calculated from the product of the velocity $v$ and assumed scattering rates $\tau(E)$, as indicated for acoustic phonon scattering, resulting in constant $mfp(E)$.}
\label{P01_02}
\end{figure}

Many books and literature references describe the MC process \cite{Jacoboni1983,Jacoboni_LugliBook1989,Hess1991,VasileskaBook2010,TomizawaBook1993,Kosina2003_2}, thus, in this work we provide only the essential details that help with the discussion of the method we present. In general, to evaluate the transport behavior of a system, a synchronous ensemble of particles (shown in Fig.~\ref{P01_01}(a)) are randomly initialized in the domain according to the material's density of states and carrier statistics. Their trajectories are simulated by alternating between free-flight and scattering events according to the intrinsic material scattering rules. The simulation is performed under a driving force, which could be an electric field, or a temperature gradient. Typically, after initialization, a time step $\Delta t$ is selected, during which a few free-flights followed by scattering events (including self-scattering \cite{LundstromBook2000}) occur. All particles' trajectories are traced in the channel material. Both left-to-right and right-to-left going particles are simulated, and the transport properties are evaluated from the difference of the fluxes of the opposite-going particles. Customarily, there are two MC methods employed in the literature - the `ensemble' and the `incident flux' MC methods\cite{LundstromBook2000,NeophytouBook2020}. In the ensemble MC, a synchronous ensemble of particles is simulated and traced simultaneously with periodic boundary conditions for particles that reach the edges of the domain (see Fig.~\ref{P01_01}(a)). In the incident flux (or single-particle) approach an adequate amount of particles are initialized at the contacts and injected and traced one by one in the domain until they exit from a contact (see Fig.~\ref{P01_02}(a) further below). In both cases the simulation stops when adequate statistics have been collected to have convergence in the conductivity. These techniques have been successfully employed in many different settings, e.g. electronic devices, thermal materials, even nanostructured materials \cite{MoglestueBook1993,Jungemann2001,Wong2011,Wolf2014_1}. In this work, we focus on electronic transport under linear response when discussing the MC specifics, however, with appropriate modifications the method we describe could apply to phonon transport as well. The method will specifically be applied here for nanostructured TE materials. \\
\textbf{MC challenges encountered in nanostructures:} The main difficulty that arises for MC methods for nanostructures, is that the presence of nanostructuring features such as built-in potential barriers, multiple back scattering events due to embedded features of grain boundaries, pores, nanoinclusions etc (see Fig.~\ref{P01_01}(b)-\ref{P01_01}(c)), to name a few, reduce the particle flux and our ability to gather adequate statistics. Furthermore, classically, potential barriers limit the number of carriers that participate in transport to the ones that can actually flow over them (see Fig.~\ref{P01_01}(b)). Figure~\ref{P01_01}(d) compares the time taken for simulating $10^4$ electrons per energy point (injecting them from the left, and ray-tracing them until they exit the domain) versus the material's channel length. We simulate 100 energy points in total, thus the overall trajectories simulated are $10^6$. Four cases are shown: i) pristine material, ii) material with grain boundaries forcing reflection of carriers with 50 \% probability, iii) a porous material with 40\% porosity, and iv) hierarchical nanostructuring with pores added in addition to grain boundaries, as in a typical situation for thermoelectric materials \cite{Taborda2016}. As nanostructuring is added in the channel, especially pores in this case, the simulation time required for the same number of particles injected in the channel increases by several orders of magnitude. In Fig.~\ref{P01_01}(d) the barchart shows the relative time $t_{sim}$ with respect to the pristine material, again indicating orders of magnitude increase in simulation time, promoting the need for more efficient MC algorithms for nanostructures.\\
\begin{figure*}
\includegraphics[height=0.17\textwidth,width=1\textwidth]{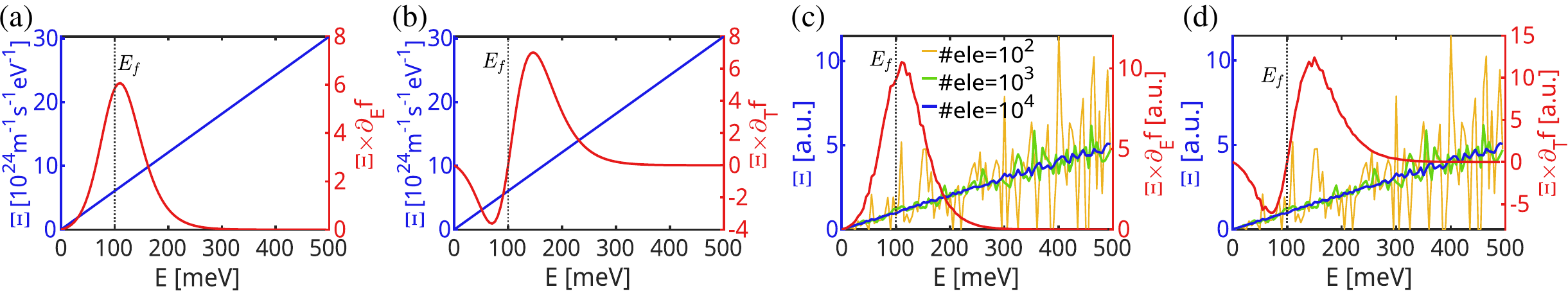}

\caption{(a) Blue line (left axis): Transport distribution function $\Xi(E)$, versus energy evaluated using the analytical Boltzmann transport equation under acoustic phonon scattering assumptions. Red line (right axis): The product of $\Xi(E)$ and differential Fermi function with respect to energy $\partial_{E} f$. (b) Same as in (a) but the right axis (red line) depicts the product of $\Xi(E)$ and differential Fermi function with respect to temperature $\partial_{T} f$. (c-d) The same as (a-b) but here the transport distribution function is calculated stochastically using the MC algorithm. The MC extracted $\Xi(E)$ is plotted for the cases of $10^2$, $10^3$, and $10^4$ simulated electrons per energy grid point. The red line (right axes) are only shown for the case of $10^4$ electrons. The right axes units of (a-b) are [$10^{25} m^{-1}s^{-1}eV^{-2}$].}
\label{P01_03}
\end{figure*}
\indent In addition, typically TE materials operate under low-field transport conditions, where the driving force is a small potential (see Fig.~\ref{P01_01}(e)), or small temperature difference at the two ends of the material \cite{Bin2017,Kasper2015,Xiaokai2015,Dirk2009,Goldsmid2014,Snyder2019,SungBook} (See Appendix C for details). This results in only a small variation between the right- and left-going fluxes around the Fermi level, which presents additional difficulties in gathering statistics for the net flux. A larger applied voltage to allow for better statistics (see Fig.~\ref{P01_01}(f)), can lead to deviation from low-field transport, and other numerical difficulties such as the need to include inelastic, in addition to elastic scattering processes, as well. In a different case, any carrier energetically positioned below the highest potential energy level (left side) will be reflected back to the right side eventually and only carriers above the source (left side) potential energy will contribute to the flux - this can also lead to difficulties in the treatment of the periodic boundary conditions. This situation becomes increasingly worse for larger applied fields. The electron-phonon scattering processes in TE materials are typically strongly influenced by elastic acoustic phonon scattering processes, and this is what we focus on in this work, although optical/inelastic processes can offer significant design opportunities \cite{Laura2019,Vassilios2019_1,Neophytou_2013,Neophytou_2019}. \\
\indent Another numerical peculiarity that is encountered in MC simulations for low energy electrons even in pristine materials, makes the computational difficulty even larger. Noticeably, the low energy carriers near the band edge have small velocities. Thus, there exists a population of slow moving electrons which ends up dominating the computational time (even so in certain cases where the Fermi level is placed higher into the bands) \cite{NeophytouBook2020}. To make things worse, in the presence of potential barriers (as shown in Fig.~\ref{P01_01}(b)), the contribution of those low-energy electrons to transport is insignificant. In addition, at these low energies, the scattering rate of ionized impurity scattering (IIS), an important mechanism especially for TEs, diverges (under low carrier densities) \cite{NagBook1980,LundstromBook2000,NeophytouBook2020}. Thus, this slow moving population of carriers could scatter very strongly, and contributes to flux even less, and thus ultimately requires unnecessarily even more computational resources. \\
\textbf{Novel MC method with improved robustness:} In this work, we develop a MC method that is more suitable for nanostructured materials, specifically addressing the difficulties described above. We use the incident flux (single-particle) approach, where the electrons are initialized at the domain boundaries one-by-one and propagate through the domain until they exit at either boundary (propagated to the other side or back-scattered - see Fig.~\ref{P01_02}(a)). \\
\indent The first issue we tackle is the large computational time associated with the ray-tracing of low energy electrons. The simulated electrons with their low group velocities end up taking a very large number of free-flight steps intervened by most probably lots of (unnecessary) self-scattering events where the electron does not scatter at the end of the free-flight as is common practice \cite{LundstromBook2000,TomizawaBook1993}. To avoid these redundant occurrences, we consider a mean-free-path ($mfp$) approach, rather than the picking of random free-flight time and the self-scattering approach. We compute the total scattering rate of the particle, and using its bandstructure velocity we calculate its mean-free-path. The particle propagates one $mfp$, and then undergoes (enforced) scattering at all times, as depicted in Fig.~\ref{P01_02}(b). In the case of acoustic phonon scattering in 3D, for example, the mean-free-path is constant in energy (as shown in Fig.~\ref{P01_02}(b)), and therefore electrons from all energies are treated in the same way, with only different free-flight durations. Since in this work we focus on introducing the new method, we only consider elastic i.e. acoustic phonon scattering process, and thus, a constant $mfp$. \\
\indent Following the incident-flux method, we use a single-injected flux from the left end of the channel, and neglect the injection of flux from the right of the channel. Thus, we refer to this from here on as the `single-flux' method, and we essentially consider only one of the two separate injections of the otherwise `two-flux' method, in which particles are injected from both left and right ends and the flux difference is computed. For these simulations we consider a two dimensional domain for numerical simplicity, with two open boundaries at the left/right, whereas the top/bottom boundaries are closed as shown in Fig.~\ref{P01_02}(a). The simulation procedure is as follows: we initialize and inject electrons in the channel domain only from the left side. We initialize those electrons uniformly in energy, rather than according to their density of states (DOS), as in typical MC methods. We typically use 1000 electrons per energy, with a uniform energy discretization steps of 5 meV. We use these for convenience, since we only consider elastic transport conditions. In the inelastic scattering case, this initialization might need to be performed according to the DOS, but it is beyond the scope of this work (such situations will be addressed in subsequent works). Then one-by-one the electrons are ray-traced, by alternating between free-flight events of a $mfp$ distance and intermediate scattering events. Here scattering only changes the particles' directions because we consider only elastic processes, thus their energy stays the same. The upper/lower closed boundaries simply specularly reflect back the electrons in the domain (surface scattering details are not considered and they are out of scope). \\
\indent To gather the flux statistics, we record the time spend in the domain by those electrons which propagate all the way from the left to the right end of the domain and exit from there (red line in Fig.~\ref{P01_02}(a)). All electrons that are back-scattered to the left do not contribute to the flux and are not considered (yellow line in Fig.~\ref{P01_02}(a)). The time an electron spends in the simulation domain until it exits to the right end is referred to as its time-of-flight ($ToF$). The average $ToF$ per particle is then computed as: 
\begin{equation}
<ToF(E)>=\frac{\sum t_r(E)}{N_r(E)}, 
\label{Eq_ToF}
\end{equation}
where, $t_r(E)$ is the time taken by a particular electron to exit from the right end, and $N_r(E)$ is the number of electrons that make it to the right end. We chose to keep the energy dependence since we only deal with elastic transport conditions. Then the $<ToF(E)>$ is used to calculate the flux per simulated electron at each energy as:
\begin{equation}
F(E)=\frac{1}{<ToF(E)>}.
\label{Eq_flux}
\end{equation}
\indent Using the flux per electron, we can form the overall flux in energy by multiplying by the density of states (DOS), $g(E)$, which essentially is proportional to the transport distribution function (TDF) of the analytical BTE as:
\begin{equation}
\Xi(E)=C\times F(E)\times g(E),
\label{Eq_Xi}
\end{equation}
This is simply because the product of flux with DOS provides essentially the flow of charge, which is directly related to conductivity in the same way the TDF determines the conductivity. \\	
\indent The proportionality constant $C$ in the equation of the TDF accounts for the super-electron charge that is typically used in MC (the fact that we only simulate a finite number of electrons), and geometrical factors related to the simulation in a finite 2D domain rather than an infinite 3D domain (and connects the conductance to conductivity). Essentially, we use $C$ to map the MC simulated TDF to the TDF that can be derived and extracted form the analytical solution of the BTE, for the case of pristine material alone. We discuss this mapping in detail further below. After having the transport distribution function numerically from MC, and calibrated to the analytical one, we can substitute it in the place of the analytical function $\Xi(E)$, which is given by $ \tau_s(E)v(E)^2g(E)$, in the usual BTE formulas below. \\
\indent The electron conductivity is calculated as:
\begin{equation}
\sigma=q^2\int_{E} \Xi(E)\left(-\frac{\partial f}{\partial E}\right)dE,
\label{Eq_sigma}
\end{equation}
and the Seebeck coefficient as:
\begin{equation}
S=\frac{qk_B}{\sigma}\int_{E} \Xi(E)\left(-\frac{\partial f}{\partial E}\right)\left(\frac{E-E_f}{k_BT}\right)dE,
\label{Eq_S}
\end{equation}
where, $q$, $k_B$ and $T$ are the electronic charge, Boltzmann constant and domain temperature (assumed constant at $T = 300 K$ throughout), respectively. The parameter $E_f$ is the Fermi energy and \textit{f} is the Fermi-Dirac distribution function. Further, the TE power factor and the electronic thermal conductivity are evaluated, respectively, as:
\begin{equation}
PF=\sigma S^2,
\label{Eq_PF}
\end{equation}
and
\begin{equation}
\kappa_{el}=\frac{1}{T}\int_{E}\Xi(E)\left(-\frac{\partial f}{\partial E}\right)(E-E_f)^2dE-\sigma S^2 T.
\label{Eq_kappa}
\end{equation}
\begin{figure*}
\includegraphics[height=0.18\textwidth,width=1\textwidth]{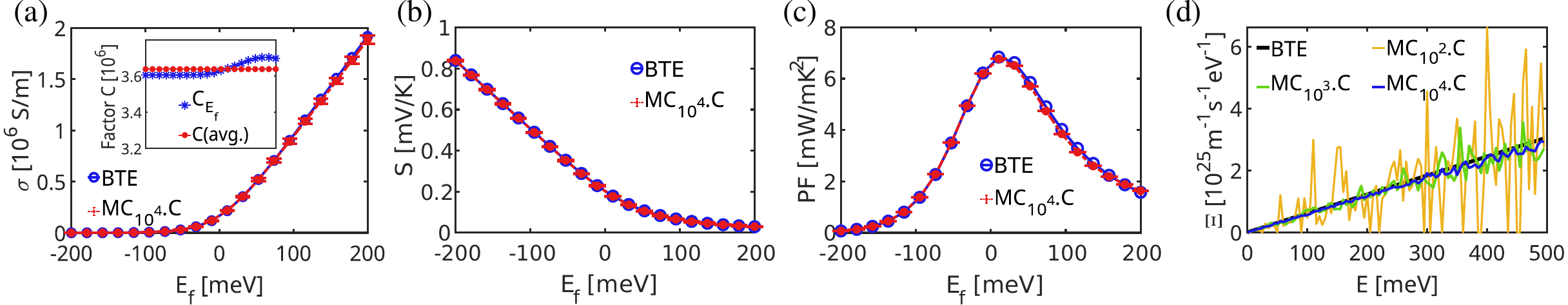}

\caption{(a) Electrical conductivity versus the Fermi energy from the analytical BTE (blue line) and the stochastic MC approach (red line). Inset: Depicts the multiplication constant that maps the MC results to the BTE results. The MC result is mapped to the BTE after after multiplication by the constant $C$. (b-c) The Seebeck coefficient and power factor for the cases in (a). (d) Transport distribution function calculated stochastically using the MC algorithm for three different electron populations per energy grid point (as indicated in the subscript of MC in the legend). The TDF maps well to the analytical one (blue line) after multiplied by the mapping constant $C$.}
\label{P01_04}
\end{figure*}
\indent The above transport coefficients are computed using MC initially for the pristine material configuration, where the calibration of the constant $C$ takes place. The idea is that once this is calibrated, we can perform MC simulations for complex nanostructured domains without further calibration and benefit form the robustness of the single-flux injection method. \\
\indent Considering 3D carriers and acoustic phonon scattering limited processes alone (which are elastic), the transport distribution function $\Xi(E)$ calculated analytically from BTE is linear in energy as shown in Figs.~\ref{P01_03}(a) and \ref{P01_03}(b) (left axes, blue lines). In order to form the integrand in Eq.~\ref{Eq_sigma} and \ref{Eq_S} above, essentially to account for linear response, we need to multiply the TDF with the differential of the Fermi distribution function with respect to energy (i.e. $\partial_E f=\frac{\partial f}{\partial E}$) for the electrical conductivity, and with respect to temperature (i.e. $\partial_T f=\frac{E-E_f}{k_BT}\times\frac{\partial f}{\partial E}$) for the Seebeck coefficient calculations. These products are shown by the red lines in Figs.~\ref{P01_03}(a-b) for an arbitrarily chosen Fermi level of $E_f$=100 meV. Using our developed single-flux MC method, we simulate the transport distribution function as well, plotted in Figs.~\ref{P01_03}(c) and (d). The linear trend of the analytical $\Xi(E)$ is also achieved in this case. However, it takes approximately $10^3$ simulated electrons per energy to gather enough statistics to match the linear trend. $10^3$ electrons per energy results in an adequate result for the transport coefficients (as we will also show below), although the TDF is still slightly noisy (green lines), while the more or less 'noise-less' $10^5$ electrons case (not shown) does not provide a noticeable conductivity accuracy advantage compared to $10^4$ (blue lines). For the case of $10^4$ electrons per energy point, we plot the $\Xi \times{\partial f_E}$ and $\Xi \times{\partial f_T}$ products in Figs.~\ref{P01_03}(c) \& \ref{P01_03}(d) (red lines) that match the trends of the analytical results as well. In this way, by obtaining the TDF from MC, and multiplying by the differential of the Fermi distribution, we effectively eliminate: i) the need for two counter propagating simulation fluxes (now this is captured by the differential functions), and ii) the need for an application of a driving force, either a voltage difference or a temperature difference, thus avoid the peculiar situation described in Figs.~\ref{P01_01}(e) \& \ref{P01_01}(f) where a small enough potential difference window does not provide enough statistics, while a large enough could make the simulation deviate from linear response or from the range of voltages that TE materials utilize (see Appendix C). Note also that the acquisition of adequate statistics is even more difficult in the case of a temperature gradient for the Seebeck coefficient in common bi-directional flux methods. We do not only need to differentiate between the right and left going fluxes, but also the ones which flow above and below the Fermi level. \\
\textbf{Mapping constant $C$:} Here we provide details on how we obtain the mapping constant $C$ between the simulated MC TDF and the analytical BTE TDF for the pristine material. For this we calculate the electrical conductivity as a function of the Fermi energy from both approaches (by integrating the TDF times the Fermi derivative in energy), and then find the mapping factor $C_{E_f}$ at each Fermi energy, essentially the ratio of the two conductivities. As we can see in the inset of Fig.~\ref{P01_04}(a), that value is almost constant for all $E_f$, so we take the average of that i.e. $C_{avg}$ as the overall final $C$, which gives only one constant factor altogether for the whole range of $E_f$ for mapping the MC conductivity to the analytical one. Since $C$ multiplies the flux, it will be used to obtain the rest of the thermoelectric transport coefficients as a function of $E_f$ as shown in Fig.~\ref{P01_04}. Clearly, we obtain an excellent match for the electrical conductivity, Seebeck coefficient, and power factor between the MC and analytical BTE as a function of Fermi energy $E_f$, as plotted in Figs.~\ref{P01_04}(a-c), respectively. The electronic thermal conductivity also matches similarly well (not shown).\\
\indent We used the overall integrated conductivity to extract the constant $C$, but it is interesting to compare how the transport distribution function extracted from the MC flux method and afterwards multiplied by the computed $C$, compares to the analytical TDF from BTE. Figure~\ref{P01_04}(d) shows this comparison, indicating excellent agreement with the analytical TDF (essentially the black line for the analytical curve resides directly below the blue line for the MC flux TDF). Even the MC lines extracted using a rather small number of simulated carriers, resulting in noisy $\Xi(E)$, also reside around the analytical line. Finally, notice the increasing variations in the flux with energy in MC, more noticeable for the low numbers of simulated electrons per energy. The oscillations in $\Xi(E)$ are stronger at higher energies, but as a matter of fact these oscillations, which are just statistical variations, appear uniformly in energy in the calculations of the $ToF(E)$ and flux $F(E)$. They are of similar relative amplitude compared to the mean values of $ToF(E)$ (or $F(E)$) in energy. For the $\Xi(E)$, however, at low energies: i) these oscillations have a smaller absolute amplitude since the flux is lower there, ii) but also they are scaled by lower density of states, compared to the high energy part of $\Xi(E)$, which further reduces their amplitude. We explain this in more detail in Appendix A. \\	
\section{Error analysis and convergence}
\begin{figure}
\includegraphics[height=0.55\textwidth,width=0.5\textwidth]{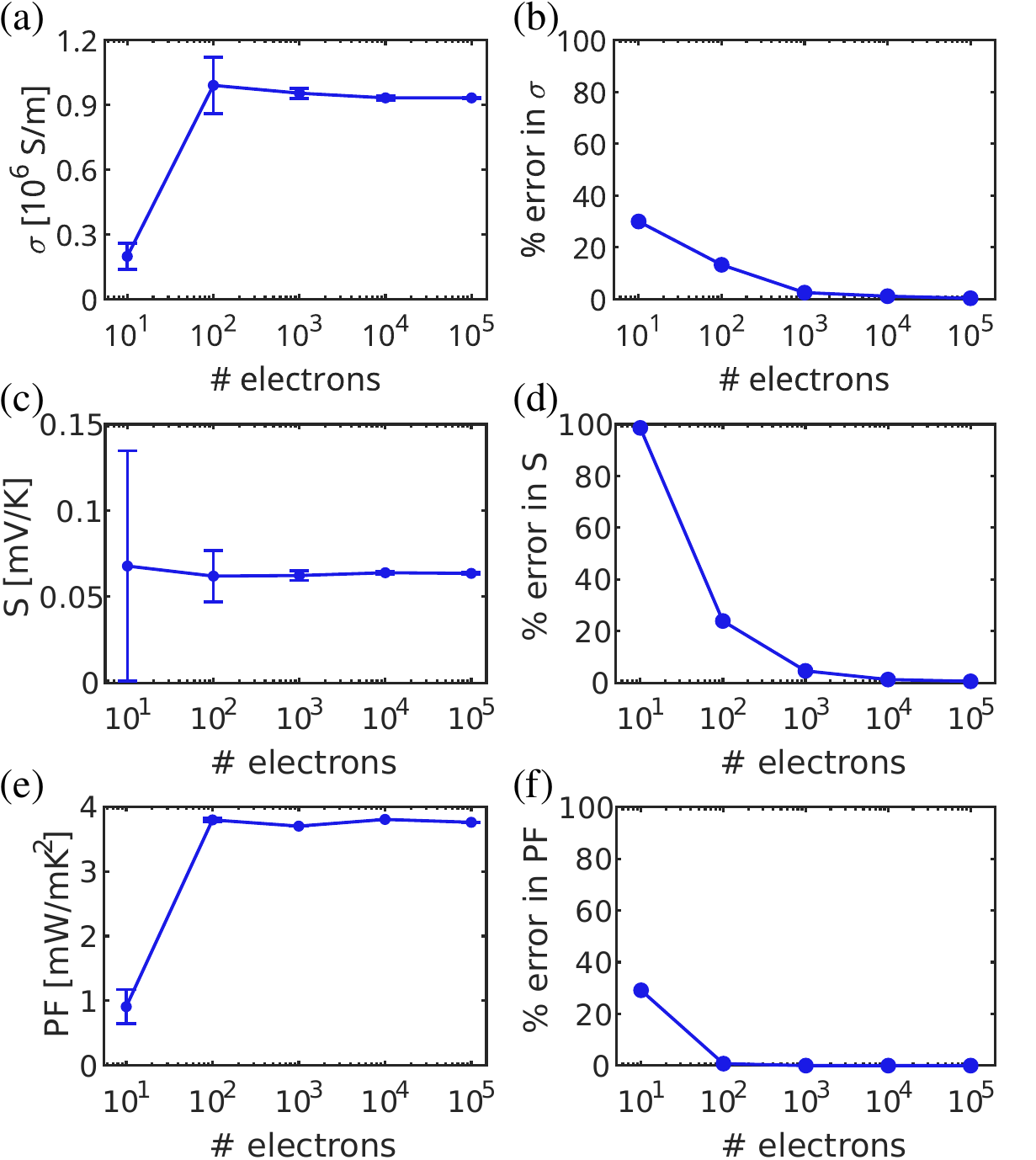}

\caption{(a, c, e) The electrical conductivity, Seebeck coefficient, and power factor, respectively, for a pristine material versus the number of simulated electrons per energy grid point. (The total number of electrons is 100$\times$ larger as we use 100 energy grid points). The stochastic variation in these quantities (error bars) after 10 repetitions of each simulation is indicated. (b, d, f) The percentage (\%) error with respect to the mean value of respective quantity for the cases in (a, c, e), respectively.}
\label{P01_05}
\end{figure}
\indent The method we present, requires only a single flux injection from one contact into the channel, does not require the application of a driving force, and bypasses the time consumed by the low energy electron ray-tracing process, all of which make it computationally very efficient. It is tested at this point for elastic scattering processes alone and further work will be required to make it compatible with inelastic processes as well.	
\begin{figure*}
\includegraphics[height=0.17\textwidth,width=1\textwidth]{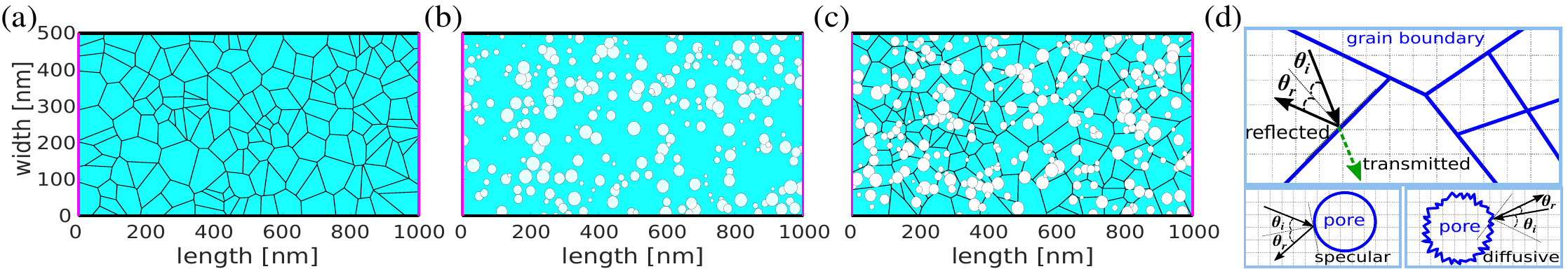}

\caption{Schematic of complex nanostructured materials in a 2-dimensional domain populated with (a) grain boundaries, (b) pores, and (c) combination of grain boundaries and pores. (d) Schematic of reflection and transmission processes (chosen randomly with probability 0.5) across a grain boundary in the domain, with the incident and reflected angles $\theta_i$ and $\theta_r$ indicated, respectively. The lower panels show reflections from a pore boundary, either under specular (used in this work) or diffusive conditions.}
\label{P01_06}
\end{figure*}
To further stress the error suppression and computational effectiveness of this single-flux method we present, in Figs.~\ref{P01_05}(a, c, e), we plot the conductivity, Seebeck coefficient and power factor with error bars versus the number of simulated electrons. We show results for 10 independent simulation runs for each data point, and the error bars are a measure of the variation within these 10 runs. In each simulation run we use $10$ up to $10^5$ electrons per energy point in increments of an order of magnitude. As mentioned earlier, we use 100 energy points in total. A single Fermi energy ($E_f$=100 meV) is used in the calculations. In Figs.~\ref{P01_05}(b, d, f) we plot the percentage standard deviation (\% error) of the corresponding transport coefficients versus the simulated electrons per energy point (error bar sizes). However, we normalize these values to the average value of their corresponding simulation run and present the percentage error. For both the conductivity and the Seebeck coefficient it appears that $10^3$ electrons per energy point (i.e. $10^5$ in total) already significantly reduce the simulation variation to a few percentage points while the transport coefficients converge to their final value. The power factor, since it composes of quantities which are inversely proportional (conductivity and Seebeck), is in general a less-fluctuating quantity and requires only $10^2$ electrons per energy to reach very close to its convergent value. Note that these simulations are executed within minutes ($\approx$ 2-3 minutes). This makes this method much more computationally efficient compared to common MC works \cite{Dhritiman2018,Wolf2014_1,TermentzidisBook2014}. For further comparison details see Appendix B.
\section{Method effectiveness for nanostructures}
\begin{figure}
\centering
\includegraphics[height=0.3\textwidth,width=0.45\textwidth]{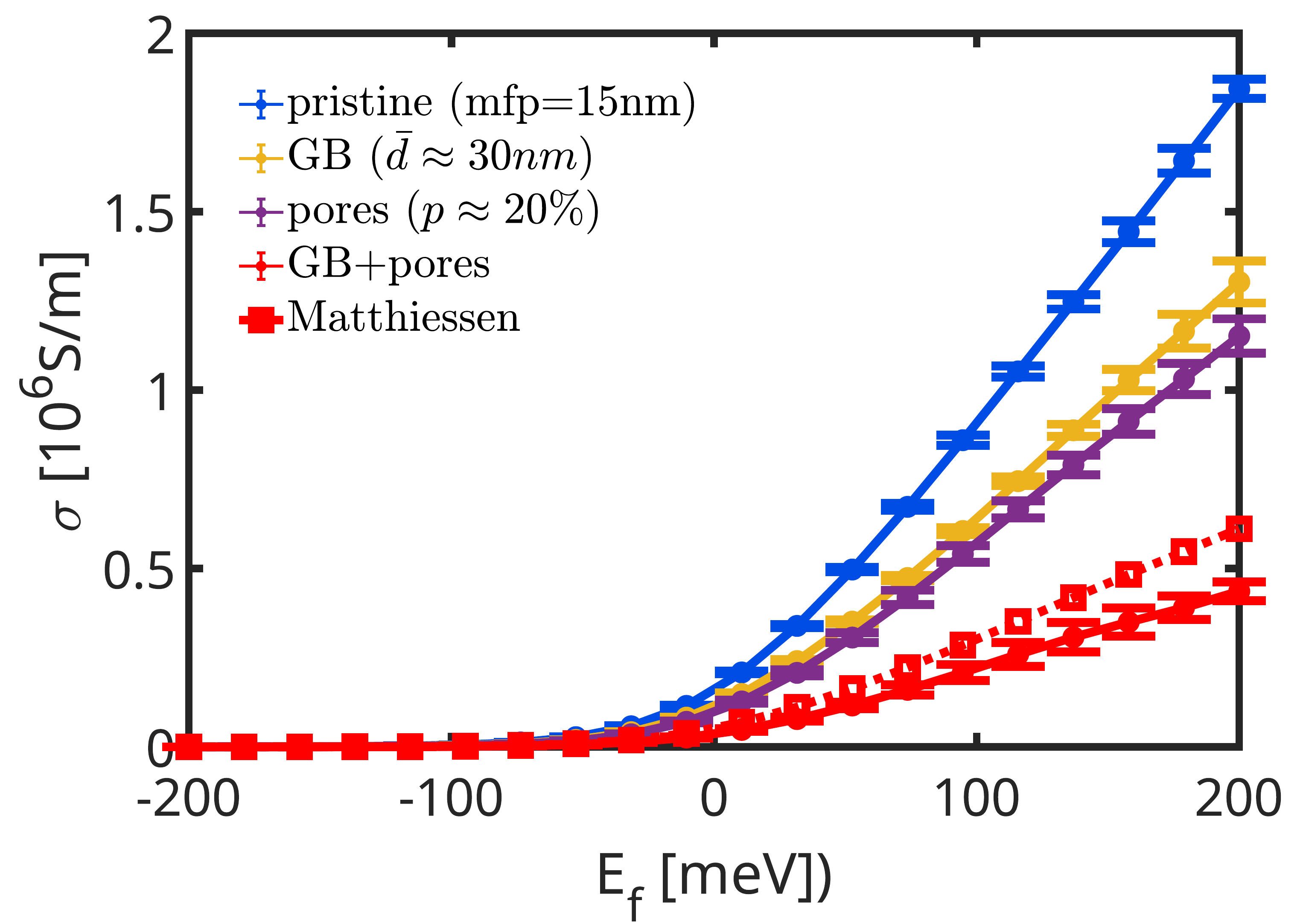}
\caption{Electrical conductivity calculated from the MC algorithm in nanostructured material domains populated with grain boundaries (yellow line), pores (purple line) and the combination of both (red line) versus the Fermi energy. The red-dotted line shows the calculated conductivity using Matthiessen's rule. The blue line indicates the pristine material conductivity.}
\label{P01_07}
\end{figure}
\begin{figure}
\includegraphics[height=0.55\textwidth,width=0.5\textwidth]{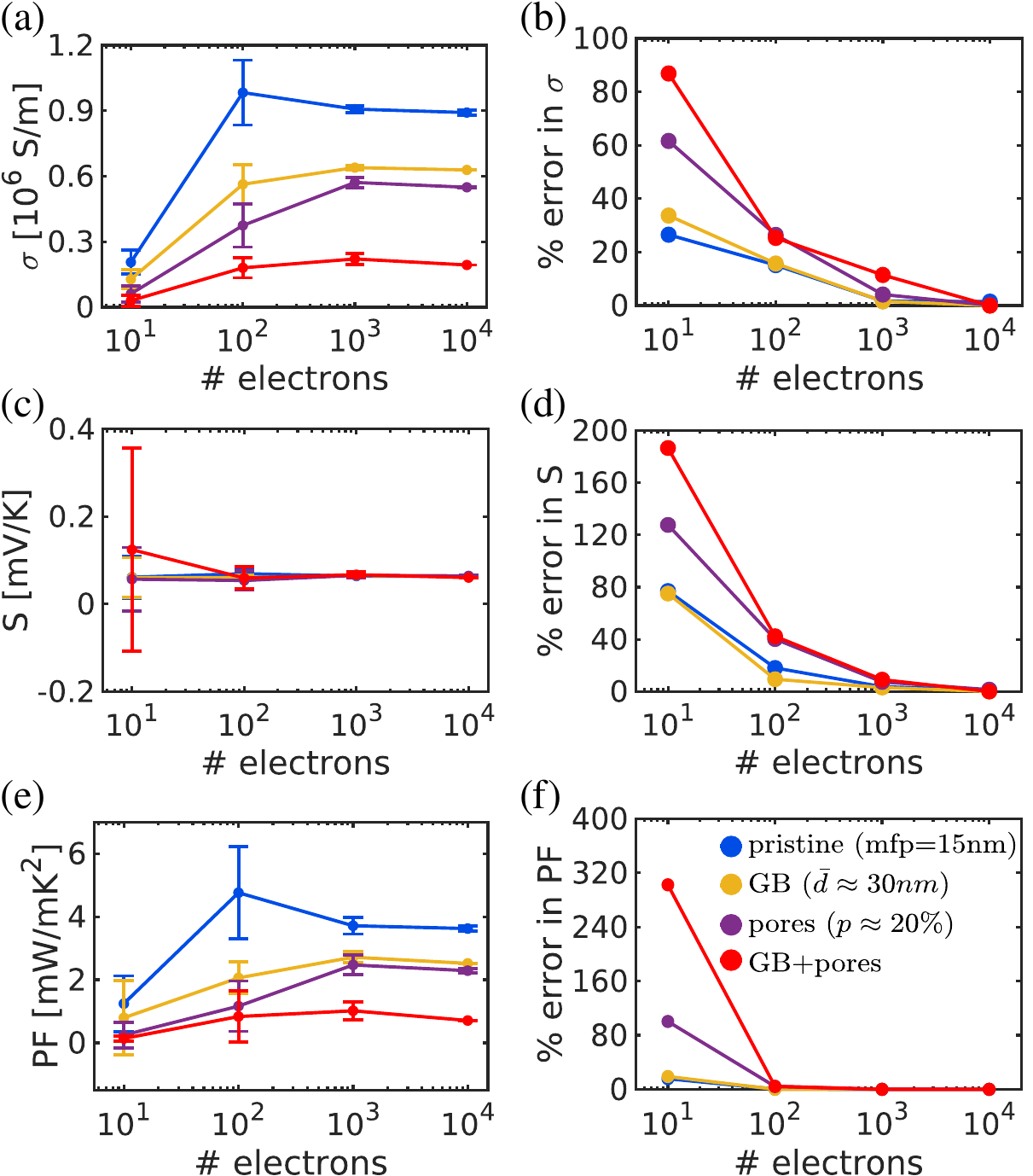}

\caption{(a, c, f) The electrical conductivity, Seebeck coefficient, and power factor, respectively, for the channel domains as shown in Figs.~\ref{P01_06}(a-c) versus the number of simulated electrons per energy grid point. (The total number of electrons is 100$\times$ larger as we use 100 energy grid points). The stochastic variation in these quantities (error bars) after 5 repetitions of each simulation are indicated. (b, d, f) The percentage (\%) error with respect to the mean value of respective quantity for the cases in (a, c, e), respectively. The plots are extracted at $E_f=100 eV$. The legend in (f) refers to all subplots.}
\label{P01_08}
\end{figure}
\indent We now demonstrate the effectiveness of the proposed single-flux method for nanostructured materials. We simulate the geometries shown in Fig.~\ref{P01_06}. First we populate the two dimensional domain with noncrystalline grain boundaries, as shown in Fig.~\ref{P01_06}(a), using the Voronoi algorithm \cite{Voronoi_MATLAB,Voronoi_Wolfram}, as implemented in Matlab. The algorithm takes seeding points and the domain dimension as inputs to create the grain boundaries. The average size of the grain is calculated by taking the mean of the length of all lines joining all nearest neighboring grain seeds by the Delaunay triangulation algorithm\cite{Voronoi_MATLAB}. During transport, when an electron encounters these grain boundaries, a random decision is made whether the electron will reflect or transmit through, as depicted in Fig.~\ref{P01_06}(d). Many detailed models exist to describe the transmission and reflection probabilities, which can be momentum, energy, and angle of incidence dependent as well \cite{Dhritiman2018,Wolf2014_2,Aksamija2014,LacroixPRB2014}. However, since here our focus is to demonstrate our new algorithm, and not the details of boundary scattering, we allow for 50\% transmission and 50\% reflection probability, independent of the carriers' energy and direction.\\
\indent For the second class of nanostructures, we create randomly distributed nanopores in the domain (with pore sizes $5<d<20$ nm) as shown in Fig.~\ref{P01_06}(b). The average porosity in the domain depends on the number of pores and their sizes. The shape of the pores can be circular, elliptical (oval), with the assumption of specular or diffusive scattering, i.e. with sharp or irregular rough edges in a realistic scenario, respectively (as shown in Fig.~\ref{P01_06}(d)). Upon scattering of a particle on a pore, the angle of incidence is computed, and the direction of reflection is determined according to the nature of the pore boundary (diffusive or specular). In this case, for simplifying the scattering process, we use specular reflection at the pores, which means that the angle of incidence is equal to the angle of reflection. We then combine these two features and make the domain hierarchically more complex as shown in Fig.~\ref{P01_06}(c).\\	
\indent In Fig.~\ref{P01_07} we show the electrical conductivity for these structures as a function of Fermi energy $E_f$. The uncertainly related to the error bars shown is a result of the MC statistical variation of simulating the same structure 5 times. The conductivity decreases when we nanostructure the domain as compared to the pristine material (blue line). There is an average drop of (i) $30\%$ in the conductivity of the channel which includes the grain boundary features, (ii) $\approx40\%$ in the case of nanoporous features, and (iii) a drastic decrease of $\approx80\%$ when both features are combined. These reduction levels are expected, considering that we have set a 15 nm mean-free-path for the pristine matrix, and the average grain size and distance between pores is of the order of 30 nm. We also find a good agreement in the MC results as compared to Matthiessen’s rule (depicted by the red-dotted line) and computed by adding the resistivity obtained at each step (as a measure of the scattering rates) as:
\begin{equation}
\frac{1}{\sigma_{Matthiessens}}=\frac{1}{\sigma_{ph+GB}}+\frac{1}{\sigma_{ph+pores}},
\label{Eq_Matthiessens}
\end{equation}
where, $\sigma_{ph+GB}$ is the conductivity due to phonon and grain boundary scattering and $\sigma_{ph+pores}$ is due to phonons and pore scattering.  \\
\indent We now examine the error in the stochastic calculation of transport coefficients versus the required number of simulated electrons for convergent results, as well as the variability due to different simulation runs (here we have executed five MC runs per data point). As shown in Figs.~\ref{P01_08}(a, c, e), the calculated quantities are well converged for $10^3$ electrons per energy grid point and with almost negligible errors (shown in Figs.~\ref{P01_08}(b, d, f)) in the variability of different simulation runs. This is an observation very similar to the pristine channel indicating that the accuracy of the method does not degrade with nanostructuring geometry complexity. Beyond $10^3$ electrons per energy grid point the variation per MC run is almost stagnant, so we used the $10^3$ carrier case for the data in Fig.~\ref{P01_07} earlier. However, more carriers can reduce the error more at the expense of more computational resources. A total of $10^5$ ray-racing trajectories is significantly smaller compared to typical works using standard MC algorithms for such structures, which require millions of trajectories. Note that these simulations with $10^5$ total trajectories take approximately 20 hrs on a single CPU to be executed. \\
\indent Experimental transport studies in nanoporous materials overwhelmingly consider mostly heat transport. We were able to identify a study with measured electronic transport data in nanoporous Si \cite{Boor2012}. The paper compared the mobility of pristine Si with 60\% porosity and reported $\sim4$ times mobility reduction for n-type samples and between $\sim3-7.5$ reduction for p-type samples. Despite the many uncertainties of the experiment, we have simulated a channel with 60\% porosity (same as in the experiment) and computed the reduction in electrical conduction compared to the pristine channel to be $\sim3.5$, which is similar to what measured in the experiment.\\ 
\indent Finally, we would like to mention a couple of things that were omitted in our implementation, that are typically present in transport codes. We have considered only acoustic phonon scattering, but including more scattering mechanisms such as by optical phonons, ionized impurities and defect scattering, would be treated in the same way as in any existing Monte Carlo method \cite{TomizawaBook1993,LundstromBook2000,VasileskaBook2010}, but a mean-free-path will be derived rather than a time-of-flight. This does not alter the performance advantage of our method, which is based on eliminating the subtraction of two very similar fluxes. We also note that in all of our simulations we used a flat conduction band profile at 0 eV. In reality, nanostructured materials have a varying conduction band profile in the channel, as a result of a varying electrostatic potential due to charging effects caused by the nanoinclusions and the grain boundaries. This is typically captured in most transport simulators by self-consistent coupling of carrier statistics (for the local charge density) with the Poisson equation (for the potential). We have not considered a varying potential in the simulation domain, but this is a separate calculation that can be performed independently prior to the MC ray-tracing part. In that case, as the particles move in the channel, their different attributes, e.g. their velocity, scattering rates, mean-free-paths, etc., would change as dictated by their energy relative to the underlying band profile. With regards to time-dependent simulations, this incident-flux method might not be the most relevant one, to the best of our knowledge.
\section{Conclusions}
\indent In summary, we have presented a novel Monte Carlo ray-tracing algorithm with computational efficiency of at least an order of magnitude compared to existing algorithms \cite{Wolf2014_1,Wolf2014_2,Dhritiman2018}. Our new method is a hybrid between the analytical Boltzmann transport and stochastic Monte Carlo, and considers many deviations from the common Monte Carlo transport algorithms which make it specifically efficient for low-field transport in nanostructured materials. For example, in our algorithm, we do not consider the timing griding as it involves several self-scattering events, which especially for the low-energy electrons it is a computationally expensive process with little influence on the results. Instead, we employ a mean-free-path approach. The method uses significantly less number of ray-tracing electrons compared to standard Monte Carlo algorithms for similar problems, avoids the statistically challenging subtraction of two opposite going fluxes, and avoids the application of driving external forces all together. We demonstrated the algorithm's efficiency and strength for accurate simulations in large nanostructured domains with multiple defects. The method is convenient for studying electronic transport in highly complex nanostructured materials, especially in the field of thermoelectrics. It is tested for elastic transport and scattering conditions, but can be extended to include inelastic processes, as well as go beyond electronic, to phonon transport (see Appendix D for a summary of how this can be implemented). We believe the new method provides an efficient and user friendly algorithm, which will enable the proper study of highly nanostructured materials under low-field steady-state conditions.\\

{\it{Acknowlegements:}} The work received funding from the European Union’s Horizon 2020 research and innovation program under grant agreement No 863222 (UncorrelaTEd). \\
\begin{appendix} 
\section{{More noise in $\Xi(E)$ at higher energies} \label{appendix_01}}
\begin{figure*}
	\centering
	\includegraphics[height=0.22\textwidth,width=0.9\textwidth]{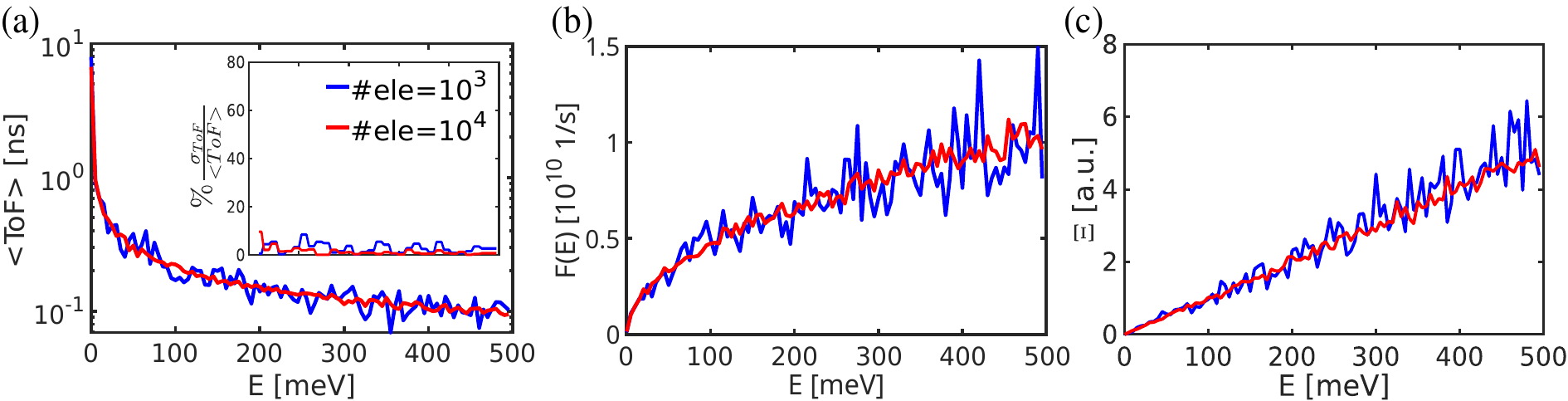}
	
	\caption{(a) Time of flight $ToF(E)$; (b) electron flux; and (c) transport distribution function $\Xi(E)$, versus energy, calculated from the Monte Carlo algorithm for two different electron simulation numbers. For the larger number of electrons, smaller fluctuations are observed (red lines). Inset of (a): the percentage deviation with respect to the mean value of the $ToF(E)$, indicating that this is overall constant in energy. The legend in (a) corresponds to all subplots in this figure.}
	\label{P01_Appendix01}
\end{figure*}
As depicted in the main text, the transport distribution function $\Xi(E)$ calculated using the developed Monte Carlo ray-tracing algorithm has more fluctuations at higher energies. In order to understand this we plot in Fig.~\ref{P01_Appendix01}(a)-(b) the energy resolved time-of-flight $ToF(E)$ and its inverse, the flux $F(E)$ for the cases of $10^3$ (blue lines) and $10^4$ (red lines) simulated electrons per energy grid point. The $ToF(E)$ follows a decreasing trend due to the higher velocities of electrons at higher energies, while $F(E)$ increases. Note that we find that the average number of electrons exited to the right is almost constant with energy in our elastic simulations. We observe statistical fluctuations around the local mean value of the $ToF(E)$ - note that we plot this in logarithmic scale. The inset plot in Fig.~\ref{P01_Appendix01}(a) shows the percentage variation of the simulated values with respect to the mean value (i.e. percentage variation of the ratio of local standard deviation to the local $ToF$ mean value). That percentage is low, but importantly is also rather constant in energy (as expected). The flux  $F(E)$ in Fig.~\ref{P01_Appendix01}(b) is plotted in linear scale, as it will enter the $\Xi(E)$ evaluation. The inverse of high $ToF(E)$ fluctuations at low energies appears now reduced compared to those at higher energies, although as we showed above, they are relatively of the same magnitude compared to their local mean value in energy. When it comes to form the $\Xi(E)$, and we multiply by the DOS, $g(E)$, the low energy fluctuations are further scaled due to the lower DOS at lower energies, but magnified at higher energies due to the larger DOS, as depicted in Fig.~\ref{P01_Appendix01}(c). For increased number of electrons the fluctuations become smaller compared to their average local value in energy as shown by the red lines. Thus, it appears that at higher energies we have the presence of more noise.\\

\section{{Reduced number of ray-tracing particles} \label{appendix_02}}
\begin{figure}
	\centering
	\includegraphics[height=0.18\textwidth,width=0.5\textwidth]{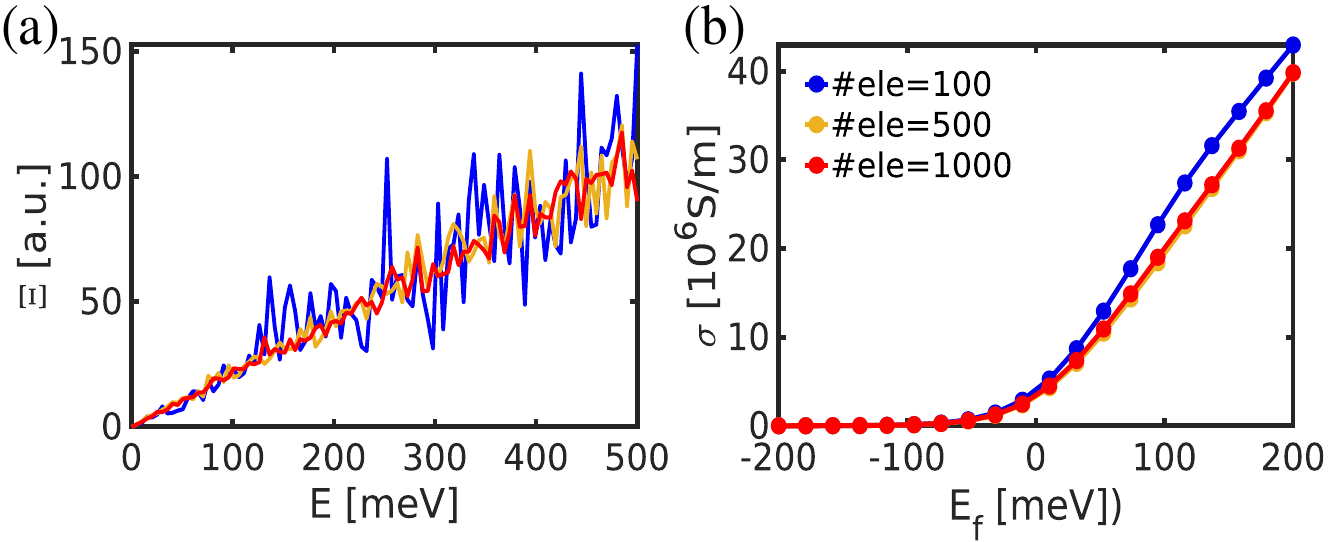}
	
	\caption{(a) Transport distribution function $\Xi(E)$ for a $200~nm$ channel length pristine domain with different number of simulated electrons per energy grid point. (b) The conductivity for this channel versus Fermi energy. The legend in (b) corresponds to both subplots in this figure.}
	\label{P01_Appendix02}
\end{figure}
In our Monte Carlo method, for a $1000~nm\times500~nm$ two-dimensional channel size, under extreme nanostructuring, the transport results converge when using approximately $10^3$ ray-tracing particles per energy. In the case of pristine material, convergent results with reduced error can be achieved for even less particles (see Fig.~\ref{P01_08}). By using 100 energy points, adequate results are thus obtained with $10^5$ trajectories in total in the nanostructured channels. The simulations we present in the paper, the length of channel and the excessive nanostructuring, are reasons why only a small percentage of particles make it from the left to the right and contribute in building the transport distribution function. In fact, only 2-3 percent make it from the left to the right, while the majority is back-scattered to the left contact, thus, a large portion of the particles do not trace the full length of the channel either, with a reduced computational cost. \\
\indent To estimate the computational benefits of this single-flux method compared to other similar Monte Carlo works, we provide below some information about the number of ray-tracing trajectories typically employed. We are not aware for Monte Carlo electronic transport works in the nanostructures we consider and the domain sizes we consider, thus, we compare first with our own prior ‘two-flux’ method in phonon transport Monte Carlo works (the ray-tracing part of the codes can still be compared). In Ref. \cite{Dhritiman2018}, on 2D phonon transport in similar nanostructured domains, 2.5 $\times 10^6$ particles were required under the `two-flux' incident flux method. Our work by Wolf et.al. \cite{Wolf2014_1,Wolf2014_2}, used more than $10^6$ ray-tracing particles for a 3D $1000~nm$ channel length again using the 'two-flux MC' incident-flux method. These are typical numbers for many MC (phonon transport) works on micrometer length scale domains, which are more than an order of magnitude larger compared to the single-flux method we present here. \\
\indent The situation is different if the channel is shorter, were more particles cross the channel and, thus, less overall need to be eventually simulated. In Fig.~\ref{P01_Appendix02}(a), we show the transport distribution function and conductivity from simulations of a pristine material of a $\approx 200 ~nm$ size channel length. Although the TDF still has some noise when we use only $10^2$ particles per energy, the error in the conductivity (see blue line in Fig.~\ref{P01_Appendix02}(b)) is small compared to the $10^3$ particle case (red line), while the $5\times10^2$ case (orange line) shows only minimal error. Indicative, these simulations for the pristine channel either for the $200~nm$ or the $1000~nm$ channel cases, are concluded within seconds to minutes. \\

\section{{Linear response and the 2-flux MC} \label{appendix_03}}
\begin{figure}
	\centering
	\includegraphics[height=0.18\textwidth,width=0.5\textwidth]{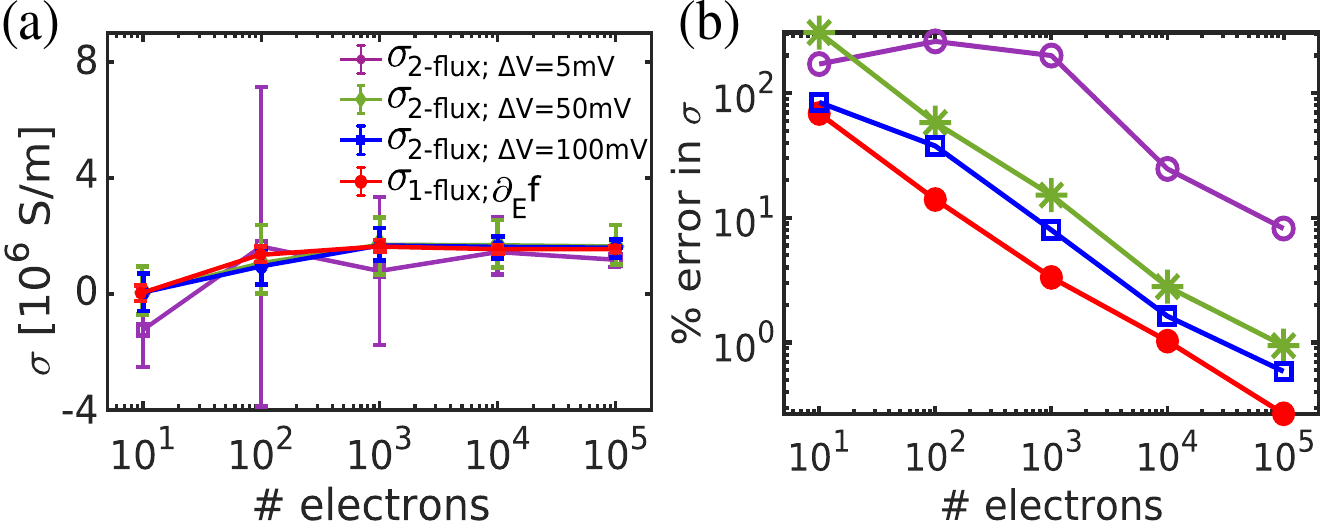}
	
	\caption{(a) Electrical conductivity with error-bars for the single-flux (red line, noted single-flux) and 2-flux MC methods versus the number of electrons simulated per energy grid point. $\Delta V$ is the biasing voltage for the 2-flux method. The calculations are performed at $E_f=100~meV$ for all cases. (b) The percentage error in the conductivities of (a) considering the stochastic nature of the single-flux and 2-flux MC methods. The legend in (a) corresponds to both subplots in this figure.}
	\label{P01_Appendix03}
\end{figure}
In common MC methods, a driving force is applied across the contacts of the simulation domain, either a voltage $\Delta V$ or temperature difference $\Delta T$. In the case of TE materials, these driving forces are small, dictating linear response within Boltzmann Transport. In a typical single TE leg simulation analysis \cite{Xiaokai2015,Dirk2009}, an n-type leg of length 10 mm develops roughly a 10 mV open circuit voltage which corresponds to 1 $\mu$V per $\mu$m voltage drop for the channels we simulate. We also mention here a work on a Bi$_2$Te$_3$ TE device \cite{Goldsmid2014}, for which the typical open circuit voltage is measured to be around 2.0 mV \cite{Snyder2019,SungBook}. Such small voltages are very hard to be simulated numerically using 2-flux incident flux or ensemble MC methods, in which the bi-directional flux difference needs to be statistically accurately resolved.
For an indication of the level of this difficulty, we simulate fluxes from the left and the right of a $1000~nm$ channel independently. For this, we place the Fermi level of the right contact lower compared to that of the left contact by $\Delta V$, essentially reducing the inflow from the right. We then obtain the conductivity by subtracting the left/right-going fluxes that were injected from the two contacts. Note that this is just an indicative scenario, since we do not apply the potential drop in the channel. For example, as $\Delta V$ increases, we should resume to the single-flux method. We compare the results with our developed single-flux method. We plot in Fig.\ref{P01_Appendix03}(a), the conductivity as a function of the number of simulated electrons for the single-flux method (red line) and for the 2-flux method under different $\Delta V$. We also include the stochastic error bar after executing 10 simulations per data point. In Fig.\ref{P01_Appendix03}(b) we plot the actual error for the different cases with $\Delta V$ as indicated in the legend of the figure. The single-flux method (red line) indicates the lowest error, which gets reduced linearly as the number of simulated particles increases. The case which uses $\Delta V=5~mV$ has the largest error, more than an order of magnitude compared to the single-flux line at the same number of simulated particles. Another way to see this (horizontally in the figure) is that 2 orders of magnitude more particles need to be simulated to reach similar convergence error compared to the single-flux method. We need to apply a larger potential difference, e.g. $\Delta V=100~mV$  as it can be seen from Fig.\ref{P01_Appendix03}(b) to reduce the error at a certain electron count number, which takes us closer to the single-flux method. However, any of these values used here are much larger compared to what thermoelectrics experience in reality. \\

\section{{Phonon transport implementation in the proposed MC algorithm} \label{appendix_04}}
\renewcommand{\thefigure}{D\arabic{figure}}

Here we provide some details on how the algorithm we present can be materialized to treat phonon transport \cite{GoldsmidBook2010,WangBook2014}. We start from the expression of the temperature derivative of the Bose-Einstein (BE) distribution (instead of the Fermi-Dirac distribution), which will provide the driving force:
\begin{equation}
	\begin{split}
		\frac{\partial f_{BE}}{\partial T} & = \frac{\partial}{\partial T} \left[ \frac{1}{exp \left( \frac{\hbar \omega}{k_BT} \right)-1} \right] \\
		& = \left[\frac{\hbar \omega}{k_BT^2}\right] \frac{exp \left(\frac{\hbar \omega}{k_BT}\right)}{\left[exp\left(\frac{\hbar \omega}{k_BT}\right)-1 \right]^2}.
	\end{split}
	\label{Eq_D1}
\end{equation}
\indent By expressing the specific heat in terms of the BE distribution derivative as well, we reach:
\begin{equation}
	\begin{split}
		C_{ph} & =\frac{\hbar^2 \omega^2}{k_BT^2}\frac{exp\left(\frac{\hbar \omega}{k_BT}\right)}{\left[exp\left(\frac{\hbar \omega}{k_BT}\right)-1\right]^2} \\
		& = (\hbar \omega)\frac{\partial f_{BE}}{\partial T}.
	\end{split}
	\label{Eq_D2}
\end{equation}
\indent A general expression for the thermal conductivity under Boltzmann transport assumptions is given by:
\begin{equation}
	\kappa=\frac{1}{2 \pi^2}\int_{\omega} \tau(\omega) v_s^2(\omega) g(\omega) C_{ph} d\omega.
	\label{Eq_D3}
\end{equation}
\indent Thus, by substituting for the specific heat, we get:
\begin{equation}
	\kappa=\frac{1}{2\pi^2} \int_{\omega} \tau(\omega) v_s^2(\omega) g(\omega) (\hbar \omega) \left[\frac{\partial f_{BE}}{\partial T}\right] d\omega.
	\label{Eq_D4}
\end{equation}
\indent Note that this expression is very similar to the one we have for the electronic conductivity, with the addition of the phonon energy term $\hbar \omega$ and the temperature derivative of the BE distribution, rather than the energy derivative of the Fermi-Dirac distribution. 
Also note that the common Callaway model can be derived from Eq.~\ref{Eq_D3} as well, following:
\begin{equation} 
	\begin{split}
		\kappa & = \frac{1}{2\pi^2} \int_{\omega}\tau(\omega)v_s^2(\omega)g(\omega)C_{ph}d\omega \\
		& = \frac{1}{2\pi^2} \int_{\omega}\tau(\omega)v_s^2 \left(\frac{\omega^2}{v_s^3}\right) (\hbar\omega) \left[\frac{\partial f_{BE}}{\partial T}\right] d\omega \\
		& = \frac{1}{2\pi^2v_s}\frac{1}{k_BT^2} \int_{\omega}\tau(\omega)(\hbar\omega)^2 \omega^2 \frac{e^{\hbar\omega/k_BT}}{\left(e^{\hbar\omega/k_BT}-1\right)^2} d\omega.
	\end{split}
	\label{Eq_D5}
\end{equation}
\indent Above we use the phonon (acoustic) density of states as $g(\omega)=\frac{\omega^2}{v_s^2}$. Now, we consider that the scattering rate for the most severe intrinsic phonon-phonon scattering event, the 3-phonon Umklapp scattering, typically follows $\tau^{-1}\sim\omega^2$. We also consider that the phonon velocity (for acoustic phonons) is constant with $\omega$. So, the frequency dependence of the phonon TDF then becomes constant with $\omega$ as $\Xi_{ph}(\omega)=\tau(\omega)v_s^2(\omega)g(\omega) \sim \text{C}$. The flux from a typical ray-tracing for a given group of particles reflects the product of mean-free-path ($mfp$) and the velocity $\sim [\tau(\omega)v_s(\omega)] v_s(\omega)\sim \text{C}/\omega^2$ (note the $mfp$ is not constant in this case with $\omega$). This is the flux simulated in the case of phonons, having a different energy dependence compared to the electronic case ($\sim\sqrt{E}$). Multiplying this by $g(\omega)(\hbar\omega)$ this time, since the thermal conductivity is the flow of energy, rather than by only $g(.)$ as in the case of electrons, we end up with again a linear dependence, $\sim \omega$, for the argument of thermal conductivity integral. Then the constant $C$ is used to map the analytical BTE model to the ray-tracing algorithm, in the same way as described for electrons in the main text. As in the case of electrons, all intrinsic material transport behavior is lumped into the simulated flux and $C$. From there on, we can focus on the effect of geometrical scattering on phonons. Thus, the formalism is very similar to the case of electrons.

\end{appendix}

\bibliographystyle{apsrev}
\bibliography{Reference}	
\end{document}